\documentclass[a4paper,11pt]{article}
\pdfoutput=1 
\usepackage{jcappub} 
\usepackage[T1]{fontenc} 
\usepackage{siunitx}
\usepackage{subcaption}
\usepackage[export]{adjustbox}
\usepackage{aas_macros}
\DeclareSIUnit \parsec {pc}

\title{Decoding a black hole metric from the interferometric pattern of the relativistic images of a compact source
}

\author[a]{Fabio Aratore}
\author[a,b]{and Valerio Bozza}

\affiliation[a]{Dipartimento di Fisica “E.R. Caianiello”, Università degli studi di Salerno, Via Giovanni Paolo II 132, I-84084 Fisciano (SA), Italy}
\affiliation[b]{Istituto Nazionale di Fisica Nucleare, Sezione di Napoli, Via Cintia, 80126 Napoli (NA), Italy}

\emailAdd{valboz@sa.infn.it}

\abstract{Photons emitted by light sources in the neighbourhood of a black hole can wind several times around it before fleeing towards the observer. For spherically symmetric black holes, two infinite sequences of images are created for any given source, asymptotically approaching the shadow border with decreasing magnitude. These sequences are reflected by a characteristic staircase structure in the complex visibility function that may be used to decode the properties of the black hole metric. Recalling the formalism of gravitational lensing in the strong deflection limit, we derive analytical formulae for the height, the width and the periodicities of the steps in the visibility as functions of the black hole parameters for the case of a single compact and distant source. With respect to diffuse emission by the whole accretion flow, this ideal framework provides clean insight and model-independent information on the metric. These basic formulae can then be used to build visibilities for more complicated sources and track the changes induced by alternative metrics and ultimately test General Relativity. As simple examples, we include visibilities for Reissner-Nordstr\"om and Janis-Newman-Winicour metrics.  
}

\keywords{black hole, gravitational lensing, interferometry}
\notoc

\begin{document}
\maketitle

\section{Introduction}
\label{intro}
The reconstruction of the first images of M87* by Very Long Baseline Interferometry (VLBI) observations performed by the Event Horizon Telescope (EHT) collaboration has represented one of the most sensational breakthroughs in present astrophysics \cite{akiyama2019first1,akiyama2019first2,akiyama2019first3,akiyama2019first4,akiyama2019first5,akiyama2019first6}. The observation of a black hole at a resolution sufficient to see its shadow surrounded by a light ring opens the possibility to directly investigate the structure of space-time in strong gravitational fields \cite{Bardeen1973} (see Ref. \cite{Perlick2021} for a comprehensive and historic review). 

The current limitations to this technique come from the locations of the antennae on the Earth surface, whose diameter provides the maximum achievable baseline \cite{akiyama2019first2}. At a wavelength of $\lambda=1.3$ mm, this corresponds to a $25 \mu as$ resolution, which is sufficient to measure the size of the shadow of a supermassive black hole in a nearby galaxy such as M87 \cite{akiyama2019first6} or the black hole in our own Milky Way, Sgr A* \cite{Zhu2019,Issaoun2019}. 

However, there is much more structure that remains hidden to current arrays that would have an extraordinary impact on our knowledge of General Relativity. The size and shape of the shadow depends on the spin of the black hole, or may be used to constrain alternative theories of gravity \cite{Zakharov2014,Tsukamoto2014,Bambi2019,Vagnozzi2019,Khodadi2020,Gralla2020,Gralla2020b,Farah2020,Shaikh2021}. The light ring surrounding the shadow can be actually decomposed in an infinite sequence of concentric rings \cite{Luminet1979} created by photons that perform one or more loops around the black hole before escaping. Such rings are exponentially thinner and dimmer as they approach the shadow border. Nevertheless, they leave an imprint on the complex visibility function, with a characteristic staircase structure \cite{johnson2020universal}. In order to see the first step of this staircase (corresponding to the resolution of the first higher order ring from the dominant outermost one), it would be necessary to build a station on the Moon or perhaps in the Lagrangian point L2. This is a strong motivation to extend the current network of arrays beyond our planet, a goal that seems in our reach in the next decades \cite{Origins2019,Radioastron2021}. Actually, what is traditionally called the ``shadow'' of the black hole is still brightened by the direct emission of the matter inflow down to the horizon. Depending on the geometry of the accretion flow, we may distinguish an ``inner shadow'' corresponding to the darkest area inside the ``shadow'' \cite{Dokuchaev2019,Dokuchaev2020a,Dokuchaev2020b,Chael2021}. Another signature of the existence of higher order images created by photons winding around the black hole would come in the time domain, since any variability in the source would be replicated an infinite number of times by higher order images with a fixed delay \cite{BozzaMancini2004}. This echos would create distinctive peaks in the autocorrelation function that could be searched for in future observations \cite{Chesler2020,Hadar2021,Wong2021}. Polarimetric images of black holes are already employed to constrain the structure of magnetic fields in the accreting matter \cite{EHTpolarizationVII2021,EHTpolarizationVIII2021} and could reveal much more with future extensions of the EHT \cite{Himwich2020,Moscibrodzka2021}. Finally, the size of the shadows of the black holes in far away galaxies may be regarded as standard rulers and help in the determination of cosmological distances \cite{Tsupko2020,Vagnozzi2020}.

The existence of relativistic higher order images has been hypothesized by Darwin back in 1959 \cite{darwin1959gravity} and recalled several times \cite{Luminet1979,Ohanian1987,virbhadra2000schwarzschild,bozza2002gravitational}. Their properties like spacing and magnification are related to the properties of the black hole metric through the coefficients of the expansion of the deflection angle in the strong deflection limit, i.e. starting from the particular trajectory corresponding to the injection of photons in the unstable circular orbit around the black hole \cite{bozza2002gravitational}. In principle, the observation of such images would allow to distinguish among alternative metrics and theories of gravity \cite{Bozza2010}. The coefficients of the strong deflection limit expansion depend on the properties of the metric at the level of the unstable circular orbit. Such properties also emerge in other contexts: the connection between strong deflection parameters and quasi-normal modes in the ringdown of gravitational waves has been widely explored
\cite{Cardoso2009,stefanov2010connection,wei2014establishing,raffaelli2016strong}.

Indeed the best chances for reaching an ultimate proof of the existence of higher order images currently come from interferometric observations at very long baselines. As shown in Ref. \cite{johnson2020universal}, higher order light rings around a black hole would generate a cascade of steps in the complex visibility that could be measured by stations in space or on the Moon. Therefore, in this paper, we propose straightforward relations between the basic properties of these steps (width, height and periodicities) and the strong deflection coefficients, thus establishing an immediate connection between such observables and the properties of the black hole metric. We will mainly focus on distant compact sources around spherically symmetric black holes, but the derived formulae can be considered as basic building blocks for more complicated source matter distributions.

The paper is organized as follows: Section 2 recalls the basics of the strong deflection limit formalism, including a discussion of extended sources. In Section 3 we calculate the complex visibility for a localized source around the black hole (e.g. a hot cloud or a stellar atmosphere, depending on the observation band) and present a gallery of pictures covering the possible geometries. In Section 4, we derive analytical formulae for the width and height of the steps created by higher order images as functions of the metric. Such formulae can be inverted to provide the coefficients of the strong deflection limit for the metric of the observed object. In Section 5, we present some examples for different metrics (Reissner-Nordstr\"om, Janis-Newman-Winicour). The source orbital motion and the time delay between consecutive images are discussed in Section 6. Section 7 contains a  discussion on the observational perspectives along with our conclusions.

\section{Generation of higher order images}
\label{Spherically symmetric black holes}

Relativistic higher order images arise around any object that is compact enough to be fully contained inside its photon sphere, i.e. the sphere covered by all unstable circular orbits of photons with any possible inclinations. So, in principle, they may exist even around objects larger than their event horizon \cite{Claudel2001,Cunha2017}. For non-spherically symmetric black holes, the unstable photon orbits at fixed radii still exist, but they do not lie on a fixed plane because of precession and their radius is a function of the angular momentum \cite{chandrasekhar1998mathematical,Teo2003}. Since our purpose is to give an analytical description of the interferometric observables in a simple case, in this paper we will focus on spherically symmetric spacetimes only and leave axially symmetric spacetimes to future works. In this framework, the general formalism for the description of gravitational lensing in the strong deflection limit has been developed in Refs. \cite{bozza2002gravitational,bozza2007strong}, which we will briefly recall in this section.

We start from the following ansatz for the line element of a generic spherically symmetric space-time

\begin{equation}
ds^2=A(r)dt^2-B(r)dr^2-C(r)\left(d\vartheta^2+\sin^2 \vartheta \ d\phi^2\right)
\label{generalsimmetricmetric}
\end{equation}

Without loss of generality, we assume orbits on the equatorial plane $\vartheta=\pi/2$. Furthermore, we assume that the function $C(r)/A(r)$ has one minimum at $r_m>0$, corresponding to the
radius of the photon sphere \cite{Claudel2001}. 

A null geodesic not ending in the black hole can be characterized through its impact parameter $u$ or through its closest approach distance $r_0>r_m$. The two quantities are related by
\begin{equation}
u=\sqrt{\frac{C(r_0)}{A(r_0)}}.
\end{equation}

It is convenient to adopt the short notation $A_m\equiv A(r_m)$ and similarly for all metric functions. In particular, a photon at impact parameter
\begin{equation}
u_m=\sqrt{\frac{C_m}{A_m}},
\end{equation}
will asymptotically approach the photon sphere and finally be injected in the unstable circular orbit.

Trajectories with a closest approach just slightly higher than the radius of the photon sphere can be described by a simple expansion

\begin{equation}
r_0=r_m(1+\delta) \qquad u=u_m(1+\epsilon),
\label{parametrizzazione}
\end{equation}
with
\begin{equation}
    \epsilon =\beta_m \delta^2 +o(\delta^2). \label{epsdelta}
\end{equation}
\begin{equation}
\beta_m=\frac{r_m^2}{4}\left(\frac{C''_m}{C_m}-\frac{A''_m}{A_m} \right).
\end{equation}

For a source at radial coordinate $r_{S}$ and an observer at $r_O$, the change in the azimutal angle $\phi$ (i.e the deflection) is

\begin{equation}
\Delta \phi=-a\log \frac{\epsilon}{\eta_O\eta_S}+b+\pi +O(\epsilon), \label{deflection}
\end{equation}
where
\begin{gather}
\eta_O=1-\frac{r_m}{r_O} \qquad \eta_S=1-\frac{r_m}{r_S}  \qquad \eta=1-\frac{r_m}{r}\label{etaS} \\
a=\frac{r_m}{u_m}\sqrt{\frac{B_m}{2A_m \beta_m}} \qquad b=a\log(4\beta_m)+b_O+b_S -\pi\\
\qquad b_O=\int_0^{\eta_O} g_1(\eta) \ d\eta \qquad b_S=\int_0^{\eta_S} g_1(\eta) \ d\eta \\
g_1(\eta)=\left[u_m\sqrt{\frac{B(r(\eta))}{C(r(\eta))}}\left(\frac{C(r(\eta))}{A(r(\eta))}-u^2\right)^{-\frac{1}{2}} \frac{r_m}{(1-\eta)^2}-\sqrt{\frac{B_m}{2\beta_m C_m}} \frac{r_m}{\sqrt{2\delta\eta+\eta^2}}\right]
\end{gather}

The integrals in $b_O$ and $b_S$ are performed on the regular function $g_1(\eta)$ and can be evaluated numerically, in general, or analytically for some simple metrics.

For a source at azimutal coordinate $\phi_S$ and an observer at $\phi_O=\pi$, we can easily find the position of the images in the observer frame by inverting Eq. (\ref{deflection}) with $\Delta \phi=\phi_O-\phi_S+ 2 n\pi$:

\begin{equation}
\epsilon_{n,\pm}=\eta_O \eta_S e^{\frac{b\pm \phi_S-2n\pi}{a}}.
\label{imagesposition}
\end{equation}

Here the double sign accounts for images appearing on the same side of the source (positive parity images) and images appearing on the opposite side (negative parity images). The latter are simply obtained by changing the sign in $\phi_S$. $n$ denotes the number of loops performed by the photon around the photon sphere before reaching the observer\footnote{Note that papers describing the light ring find more convenient to label images by the number of half-orbits \cite{johnson2020universal,Chael2021}. In our treatment, the number of half-orbits will be given by $2n$ for positive parity images and $2n+1$ for negative parity images}. With $n=0$ and positive parity we would have the direct image, which is not described by the strong deflection limit expansion. All higher order images ($n\geq 1$) will be accurately described by Eq. (\ref{imagesposition}) and Eq. (\ref{parametrizzazione}). The negative parity image of order $n=0$ is also acceptably described by the strong deflection limit if the source is on the same side of the observer \cite{Bozza2010}. For an observer in the asymptotic region, which is the most physically interesting case, the angular separation between the direction of arrival of the photon and the direction of the black hole is simply $\theta=u/r_{O}$. So we have

\begin{equation}
\theta_{n,\pm}=\frac{u_m(1+\epsilon_{n,\pm})}{r_{O}}=\theta_m(1+\epsilon_{n,\pm})
\label{angularseparation}
\end{equation}

$\theta_m=u_m/r_O$ is sometimes called the ``angular radius of the shadow of the black hole'': since all lensed images (except the direct image) of sources outside the photon sphere reach the observer from angles $\theta>\theta_m$ and the region within the angular radius $\theta_m$ should appear as almost empty \cite{falcke1999viewing}, save for foreground sources. Of course, the possible appearance of a shadow depends on the shape, orientation and velocity of the infalling distribution of matter. Gravitomagnetic simulations favor an emitting region extending down to the horizon, well below the photon sphere. Such configuration would rather produce an inner shadow surrounded by a light ring with radius $\theta_m$ \cite{Dokuchaev2019,Dokuchaev2020a,Dokuchaev2020b,Chael2021}. For this reason, recent papers prefer to refer to $\theta_m$ as a ``critical curve'' \cite{Gralla2019}. However, the wording ``critical curve'' already has a well-established mathematical meaning in more than sixty years of gravitational lensing literature. In a gravitational lensing problem in which the source moves on a fixed surface, critical curves are the loci in which the Jacobian of the lens map vanishes and where degenerate critical images appear \cite{SEF}. In our framework, when the source is perfectly aligned behind or in front of the black hole ($\phi_S=0,\pi$), the images degenerate in infinite sequences of concentric critical curves (the Einstein rings) asymptotically approaching the shadow border, which is not a critical curve itself, rigorously speaking \cite{Luminet1979}.  Although the circle with radius $\theta_m$ may coincide with a particular critical curve after an appropriate choice of the source surface (a sphere almost coinciding with the photon sphere), in general, it is not a critical curve according to the traditional gravitational lensing terminology. In the following, for simplicity we will keep referring to $\theta_m$ as the radius of the shadow, but the reader should be aware that the darkest area (if any) can show up well within this radius in realistic inflows.  

Coming back to Eq. (\ref{angularseparation}), we see that all higher order images will appear exponentially closer to $\theta_m$ as the number of loops $n$ is increased. It is also important to note that the strong deflection limit formulae presented here (Eqs. \ref{etaS}-\ref{angularseparation}) hold even for sources inside the photon sphere. In such case, $\eta_S<0$ and then $\epsilon<0$, which means that images form close to the shadow border from the inner side \cite{bozza2007strong}. Of course, there is no closest approach distance $r_0$ for sources inside the photon sphere since light rays only come outward from the source to the observer. Indeed, $\epsilon<0$ implies a complex $\delta$ from Eq. (\ref{epsdelta}), but otherwise all formulae remain perfectly consistent. Therefore, our treatment also covers sources between the horizon and the photon sphere, provided we exclude the weakly lensed direct image.

Relativistic higher order images ($n\geq 1$) appear as extremely thin tangentially elongated arcs \cite{Bozza2005}. The tangential extension of the source remains unaffected by gravitational lensing: so for a source extending by $\Delta\vartheta_S$ in the meridional direction\footnote{We remind that we have oriented our coordinates in such a way that the source, the observer and the origin lie on the equatorial plane. The tangential extension of the images is then described by the polar angle $\vartheta$ in the coordinates (\ref{generalsimmetricmetric}).}, each image will have the same extension $\Delta\vartheta_I=\Delta \vartheta_S$. The radial thickness of the image is obtained by differentiation of Eq. (\ref{angularseparation}) 
\begin{equation}
    \Delta\theta_{n,\pm}=\Delta\theta_{0} \; \epsilon_{n,\pm} \label{dthetan}
\end{equation}
\begin{equation}
    \Delta\theta_{0}=\frac{\theta_m}{a} \Delta\phi_S +\theta_m \frac{r_m}{r_S^2 } \left(\frac{1}{\eta_S}+\frac{g_1(\eta_S)}{a} \right)\Delta r_S. \label{dtheta0}
\end{equation}

The geometric factor $\Delta \theta_0$ contains information on the source distance and geometry, while each image is exponentially suppressed by $\epsilon_{n,\pm}$.

As we shall soon see, the size of the source matters in interferometric experiments, since it determines the angular frequencies at which all structures appear. In the following, we will consider a source with azimutal size $\Delta\phi_S$, polar size $\Delta \vartheta_S$ and radial extension $\Delta r_S$ and express our results in terms of $\Delta\theta_{n,\pm}$, as given by Eq. (\ref{dthetan}). 

\begin{figure}
\includegraphics[width=17cm, right]{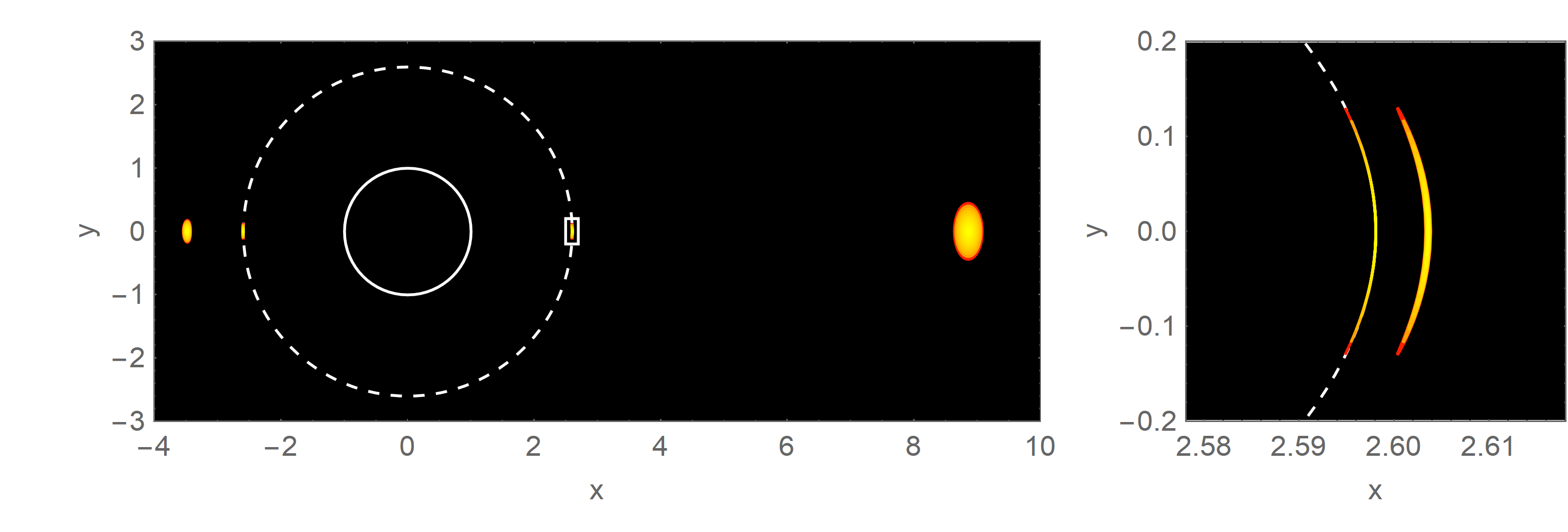}
\caption{Images generated by a source at $\phi_S=45^\circ$ at a distance of $r_S=10$ with a size of $0.5$ in units of the Schwarzschild radius $R_{Sch}=2MG/c^2$. The right panel shows a closer view of the images of order 1 and 2 with the horizontal axis stretched by a factor 10 with respect to the vertical axis. The solid line is the event horizon and the dashed line is the shadow border. The angular coordinates $x$ and $y$ are in units of the Schwarzschild radius divided by the observer distance $R_{Sch}/r_O$.}
\label{figimages}
\end{figure}

Fig. \ref{figimages} shows an example of the images generated by a static source around a Schwarzschild black hole
\begin{equation}
A(r)=c^2-\frac{2 M G}{ r} \qquad B(r)=\left(1-\frac{2 M G}{c^2 r}\right)^{-1} \qquad C(r)=r^2. \label{Schwarzschild}
\end{equation}
The major image on the right and the negative parity image on the left are the zero order images generated by relatively weak deflection. Higher order images appear right on the shadow border on both sides. The right panel shows a closer view in which the first two orders appear as separated.

The strong deflection limit formalism establishes an easy analytical framework to study the higher order images created by black holes or alternative spherically symmetric metrics. As we shall see, this will allow us to derive interesting relations between the properties of the interferometric signal and the coefficients of the metric.

\section{Interferometric signature}
\label{Interferometric signature}
Observing stations distributed around the Earth or in space \cite{akiyama2019first2,Radioastron2021,Origins2019} would probe the complex visibility function $V(\vec{u})$ corresponding to the Fourier transform of the sky image $I(\vec{x})$. Here and in the following, $\vec{x}\equiv (x,y)$ represents angular (Bardeen) coordinates in the observer sky, while $\vec{u}$ is the angular frequency vector, given by the projection of the baseline orthogonal to the line of sight and measured in units of the observation wavelength $\lambda$.

Since we are assuming the spherical symmetry for our spacetime, all images lie on the same line, say the $x$-axis. It is important to note that this is strictly true only for perfectly static sources. As the source moves, higher order images will follow with some delay, since they are generated by photons taking longer paths including $n$ revolutions around the black hole. For this reason, the description we are going to develop in this section  applies to distant enough sources, whose orbital motion is negligible compared to the time delay of higher order images. We will come back to this important issue in Section \ref{Sec alignment}. We  assume for simplicity that the source has a Gaussian profile, so that, in a first approximation, all positive parity images have a Gaussian profile with the maximum in $\vec x = (\theta_{n,+},0)$, radial size $\Delta\theta_{n,+}$ and tangential extension $\theta_{n,+} \Delta\vartheta_S$ (the tangential direction would be parallel to the $y$-axis in our coordinates, see Fig. \ref{figimages}). Negative parity images are on the opposite side, with $\vec x = (-\theta_{n,-},0)$.

Thus, the contribution of the higher order images to the sky image can be written in the following way:

\begin{equation}
I(x,y)=I_0 \sum_{p=\pm 1} \sum_{n=1}^{+\infty}  e^{-\frac{(x-p \theta_{n,p})^2}{2\Delta\theta_{n,p}^2}} e^{-\frac{y^2}{2\theta_{n,p}^2 \Delta\vartheta_S^2}}.
\label{skyimage}
\end{equation}
Note that the surface brightness is conserved by gravitational lensing. Therefore, the central brightness $I_0$ is the same for all images. Since $\theta_n =\theta_m + O(\epsilon_n)$ while $\Delta\theta_n = O (\epsilon_n)$, the extension along $x$ decays very quickly, while the extension along $y$ rapidly tends to a constant.

We can compute the Fourier transform on $I(x,y)$ to obtain the complex visibility function

\begin{equation}
V(u,v)=\int I(x,y) e^{-2\pi \imath (ux+vy)} \ dx \ dy 
\end{equation}

Each image is a product of two independent Gaussians (as long as the source is not too large). The integration simply gives

\begin{equation}
V(u,v)=2\pi I_0 \sum_{p=\pm 1} \sum_{n=1}^{+\infty} \theta_{n,p} \Delta\vartheta_S \Delta \theta_{n,p} \ e^{-2\pi^2 \Delta\theta_{n,p}^2 u^2} e^{-2\pi^2 \theta_{n,p}^2 \Delta\vartheta_S^2 v^2} e^{-2\pi \imath p\theta_{n,p} u} 
\label{visibility}
\end{equation}

The visibility function contains a real exponential, giving the decay at high angular frequencies, and a complex exponential. Note that since the zero order images are much more extended than the higher order images, their contribution to the visibility decays on much shorter scales than those described by Eq. (\ref{visibility}). In practice, while the zero order images become negligible at $u> r_O/(r_S\Delta \phi_S)$, the first order image decays at $u>1/\Delta \theta_1 \sim 1/(\theta_m \epsilon_1 \Delta \phi_S)$. In general, $\theta_m< r_s/r_O$ and $\epsilon_1 \ll 1$. Therefore, we proceed by neglecting the contribution of zero order images to very high angular frequencies.

Similarly, as the contributions due to first order images decay, the second order images stand up and so on. The structure of the visibility function in the $u$ direction is staircase-like \cite{johnson2020universal}.

The complex exponentials do not affect the modulus of the complex visibility if there is only one dominant term. However, as soon as the contribution of a dominant image decays, the complex exponential of the uprising image mixes with the previous one creating oscillations.

Eq. (\ref{visibility}) contains terms of different order in $\epsilon_n$. In order to be consistent with the strong deflection limit framework, we must only consider the lowest order terms in $V(u,v)$. With this prescription, we get some further simplification

\begin{equation}
V(u,v)=N(v)\sum_{p=\pm 1} \sum_{n=1}^{+\infty} \epsilon_{n,p} \ e^{-2\pi^2 \Delta\theta_{n,p}^2 u^2}  e^{-2\pi \imath p \theta_{n,p} u},
\label{visibilitysimple}
\end{equation}
\begin{equation}
    N(v)=2\pi I_0 \theta_m \Delta \theta_{0} \Delta\vartheta_S e^{-2\pi^2 \theta_m^2 \Delta\vartheta_S^2 v^2}.
\end{equation}

Once more, we note that the $v$ dependence orthogonal to the line on which images are formed factors out of the sum, since all images have the same tangential extension. Therefore, we have included this term in a pre-factor $N(v)$ that is the same for all terms in the sum and depends on the geometry of the source: its size, shape, brightness and position.  The $u$ dependence is much more interesting, since it contains the details of the higher order images through the amplitudes and the scales of the Gaussians. In the complex phase factor we have kept first order terms, since they become important in phase differences.

In order to visualize the visibility as a function of the angular frequency $u$, once more we consider a simple Schwarzschild metric (\ref{Schwarzschild}). For parameters like the mass or the distance between the lens and the observer we refer to $M87^*$, already studied in Ref. \cite{johnson2020universal}: $M=6.5*10^9 M_\odot$ and $r_O=\SI{16.8}{\mega\parsec}$; we take a source at a distance of $r_S=100$ with a size of $5$ in units of the Schwarzschild radius $R_{Sch}=2MG/c^2$. This could be a hot cloud resulting from the disruption of a star, or anything else coming too close to the supermassive black hole. With these values, we get $\Delta \theta_0=\SI{4.9e-12}{}$.

\begin{figure}
\centering
\includegraphics[width=\textwidth]{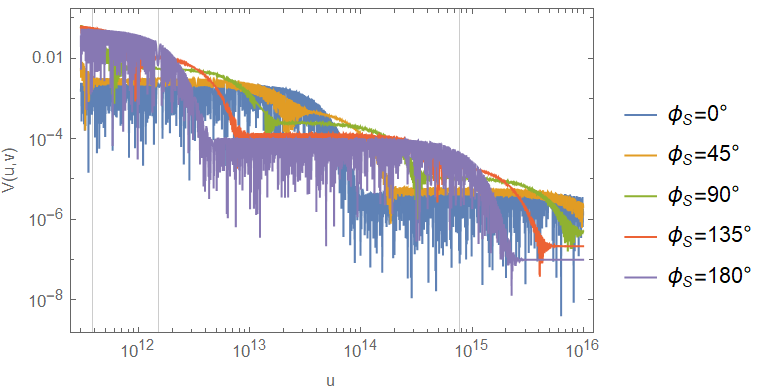}
\caption{Visibility function in Schwarzschild space-time for various source azimutal angles. The three vertical lines correspond to Moon baseline, Lagrangian point L2 baseline and Jupiter semi-major axis respectively at the wavelength $\lambda = \SI{1}{\milli\meter}$.}
\label{VcambiandoPhiS}
\end{figure}

In Fig. \ref{VcambiandoPhiS} we report the contributions to the visibility coming from higher order images. For reference, the contribution of the direct zero order image would be already negligible for $u>10^{10}$. The visibility has a staircase shape: the highest step (on the left) is due to the $n=0$ image with negative parity, the second step is the primary image at order $n=1$, the third step is the secondary image for $n=1$ and so on. Depending on the source alignment $\phi_S$, the relative positions of the positive parity and negative parity steps changes. For $\phi_S=0$ (source perfectly aligned behind the black hole, the two steps coincide (blue curve). For $\phi_S=90^\circ$, the negative parity steps are half-way between two consecutive positive parity steps. For sources in front of the black hole ($\phi_S=180^\circ$), negative parity steps coincide with positive parity steps of the following order. When two contributions are of the same order of magnitude, oscillations caused by the complex exponentials can be seen. 

\begin{figure}
\centering
\includegraphics[width=\linewidth]{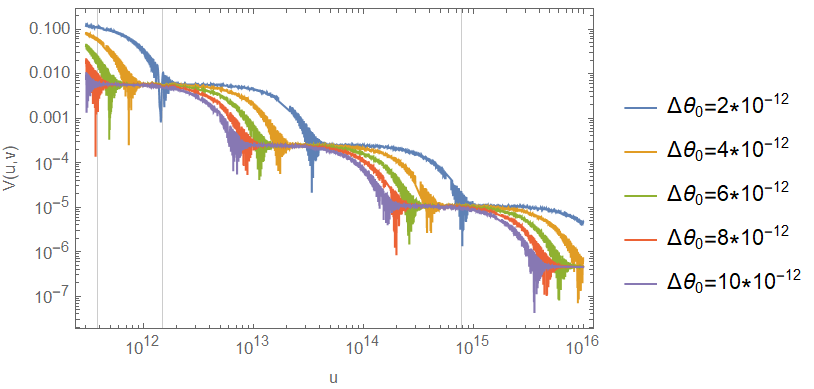}
\caption{Visibility function in Schwarzschild space-time for different values of the geometric factor with a fixed value of $\phi_S=90^\circ$.}
\label{VcambiandoDTheta0}
\end{figure}

In Fig. \ref{VcambiandoDTheta0} we see the change in the visibility with the geometric factor $\Delta\theta_0$, which depends on the source size and distance. Increasing $\Delta\theta_0$, the steps become shorter.

\section{The inversion problem: from the visibility to the black hole parameters}
\label{Width and height of each step}

Eq. (\ref{visibilitysimple}) provides a simple guidance for the visibility function generated by a compact source around a black hole or, in general, any objects endowed with a photon sphere. As noted before, the pre-factor $N(v)$ depends on the characteristics of the source, but we can easy get rid of it by taking ratios of the contributions of images of different order.

Fig. \ref{VcambiandoPhiS} shows that the visibility has a staircase structure, with each step dominated by one term in the sum in Eq. (\ref{visibilitysimple}). The height of each step is simply given by the amplitude
\begin{equation}
    h_{n,\pm}=N(v) \epsilon_{n,\pm}. \label{height}
\end{equation}

Independently of the shape of the source, the rate at which the staircase falls down only depends on the source position and the characteristics of the metric contained in $\epsilon_{n,\pm}$, as given by Eq. (\ref{imagesposition}).

We can also obtain the intersection point of each step with the following one by equating the contributions of two consecutive terms in Eq. (\ref{visibilitysimple}). We must distinguish two cases: intersection between primary and secondary image of the same order (same $n$) and intersection between secondary and subsequent primary image.

As regard the first case, we can write

\begin{equation}
\epsilon_{n,+} e^{-2\pi^2 \Delta\theta_0^2 \epsilon_{n,+}^2 u^2}=\epsilon_{n,-} e^{-2\pi^2 \Delta\theta_0^2 \epsilon_{n,-}^2 u^2}.
\label{primariavssecondaria}
\end{equation}

By inspection of Eq. (\ref{imagesposition}), we now note that
\begin{equation}
\epsilon_{n,-}=\epsilon_{n,+}  e^{-\frac{2\phi_S}{a}}.
\end{equation}

Inserting this relation in \eqref{primariavssecondaria} we can easily solve for $u$ and obtain the starting point of the step dominated by the secondary image of order $n$ with height $h_{n,-}$:

\begin{equation}
u_{n,-}=\frac{1}{\pi \Delta \theta_0 \epsilon_{n,-}}  \sqrt{\frac{\frac{\phi_S}{a}}{e^{\frac{4\phi_S}{a}}-1}}.
\label{intersezione12}
\end{equation}
Note that this formula converges to a constant as $\phi_S \rightarrow 0$. However, in the perfect alignment limit the two images merge in an Einstein ring and generate the same step save for the phase factor. 

For the second case (step between a secondary image and the primary image of the subsequent order), we write
\begin{equation}
\epsilon_{n,-} e^{-2\pi^2 \Delta\theta_0^2 \epsilon_{n,-}^2 u^2}=\epsilon_{n+1,+} e^{-2\pi^2 \Delta\theta_0^2 \epsilon_{n+1,+}^2 u^2}.
\label{secondariavsprimaria}
\end{equation}

Noting that
\begin{equation}
\epsilon_{n+1,+}=\epsilon_{n,-}  e^{-\frac{2\pi-2\phi_S}{a}},
\end{equation}
we find the starting point of the step dominated by the positive parity image of order $n$: 

\begin{equation}
u_{n,+}=\frac{1}{\pi \Delta \theta_0 \epsilon_{n,+}}  \sqrt{\frac{\frac{\pi-\phi_S}{a}}{e^{\frac{4(\pi-\phi_S)}{a}}-1}}.
\label{intersezione21}
\end{equation}

The last information we can exploit in Eq. (\ref{visibilitysimple}) is stored in the phase factor. When we calculate the visibility amplitude, the phase factors of terms having similar size create interference terms with argument $2\pi u (\theta_{n,+}+\theta_{n,-})$ or $2\pi u (\theta_{n,-}+\theta_{n+1,+})$. Such terms arise because the two images contributing at the intersection of two steps are on opposite sides of the black hole at a separation given by 
$(\theta_{n,+}+\theta_{n,-})$ or $(\theta_{n,-}+\theta_{n+1,+})$ respectively. The corresponding modulations in the visibility have a periodicities given by
\begin{equation}
    P_{n,+}\equiv \nu_{n,+}^{-1} =\left[\theta_m(2+\epsilon_{n,+}+\epsilon_{n,-})\right]^{-1} \label{Period+}
\end{equation}
\begin{equation}
    P_{n,-}\equiv \nu_{n,-}^{-1} =\left[\theta_m(2+\epsilon_{n,-}+\epsilon_{n+1,+})\right]^{-1}. \label{Period-}
\end{equation}

At this point, we can try to combine the available information to extract all possible physical parameters for our lensing problem. Indeed, the easiest quantity to measure is the  strong deflection coefficient\footnote{This is also identified as the Lyapunov exponent for the instability of null geodesics \cite{Mashoon1985,Cardoso2009}.} $a$. In fact, taking the ratio of the heights of two steps corresponding to the same parity and consecutive orders, we have
\begin{equation}
    a=\frac{2\pi}{\ln \left(h_{n,+}/h_{n+1,+} \right)}. \label{aequation}
\end{equation}

In practice, the coefficient $a$ regulates the decay rate of the visibility function by establishing the ratio of the magnification factor of images of consecutive order \cite{bozza2002gravitational}.

The next variable we can easily extract is the source azimutal position $\phi_S$:
\begin{equation}
    \phi_S=\frac{a}{2}{\ln \left(h_{n,+}/h_{n,-} \right)}.
\end{equation}

In fact, the steps corresponding to negative parity images would coincide with positive parity steps for well-aligned sources, while will come half-way between two steps of consecutive order if the source is in quadrature ($\phi_S=\pi/2$).

In order to extract the other strong deflection coefficient $b$, we need to resort to the periodicities of the oscillations from Eqs. (\ref{Period+}) and (\ref{Period-}). If we are able to measure the periods of the oscillations at two consecutive step crossings, we may substitute Eqs. (\ref{angularseparation}) and (\ref{imagesposition}), and invert 
\begin{equation}
    \theta_m=\frac{1}{4} \left\{e^{\frac{2\pi}{a}}\left[\nu_{n,-}+ e^{-\frac{2\phi_S}{a}} \left(\nu_{n,-} -\nu_{n,+} \right)\right] -\nu_{n,+} \right\}\left(\coth{\frac{\pi}{a}} -1 \right),
\end{equation}

\begin{equation}
    \eta_O\eta_S e^{\frac{b-2n\pi}{a}}= \frac{2(\nu_{n,+}-\nu_{n,-})}{\nu_{n,-}\left( e^{\frac{\phi_S}{a}} + e^{-\frac{\phi_S}{a}}\right)- \nu_{n,+} \left( e^{\frac{\phi_S-2\pi}{a}} + e^{-\frac{\phi_S}{a}}\right)}. \label{bequation}
\end{equation}

These formulae complete the full extraction of all parameters intervening in gravitational lensing in the strong deflection limit from the study of the complex visibility of one compact source around a massive compact object. Interestingly, the radius of the shadow $\theta_m$ can be obtained from the observation of a compact source, even without the observation of a full ring \cite{akiyama2019first1}. The left hand side of Eq. (\ref{bequation}) actually contains a combination of the coefficient $b$ with other geometric parameters. Assuming we know which order $n$ we are measuring, $\eta_O$ is equal to 1 for every practical observations, since we are at a distance much larger than the Schwarzschild radius. $\eta_S$ depends on the source distance through Eq. (\ref{etaS}). As stated before, we are already assuming that the source is distant enough to neglect its orbital motion. Therefore, we can also consider $\eta_S\simeq 1$.

We have not used the intersection points between the steps from Eqs. (\ref{intersezione12}), (\ref{intersezione21}) to derive the coefficients of the metric. Indeed, $u_{n,\pm}$ depends on the source size through $\Delta \theta_0$, so it can be used to investigate the nature of the source rather than the properties of the black hole. Nevertheless, it must be kept in mind that the size of the source is of critical importance to determine the necessary baseline for this kind of investigation, since all the scaling of the staircase structure depends on it: a larger source will require a shorter baseline.

\section{Examples}
\label{Results}

Eqs. (\ref{aequation})-(\ref{bequation}) provide a simple analytical framework to translate the main observable features of the complex visibility in the strong deflection limit parameters, which are connected with the underlying spacetime metric. In this section, we provide two examples that demonstrate how the characteristics of the visibility depend on the deviations from a pure Schwarzschild metric, enabling the possibility to test General Relativity with interferometric observations at an unprecedented level.

The first metric we consider is the Reissner-Nordström (RN) metric, representing the spacetime for an electrically charged static black hole

\begin{equation}
A(r)=c^2-\frac{2 M G}{ r}+\frac{q^2 G}{4 \pi \epsilon_0 c^2 r^2} \quad B(r)=\left(1-\frac{2 M G}{c^2 r}+\frac{q^2 G}{4 \pi \epsilon_0 c^4 r^2}\right)^{-1} \quad C(r)=r^2.
\end{equation}

\begin{figure}
\centering
\includegraphics[width=\linewidth]{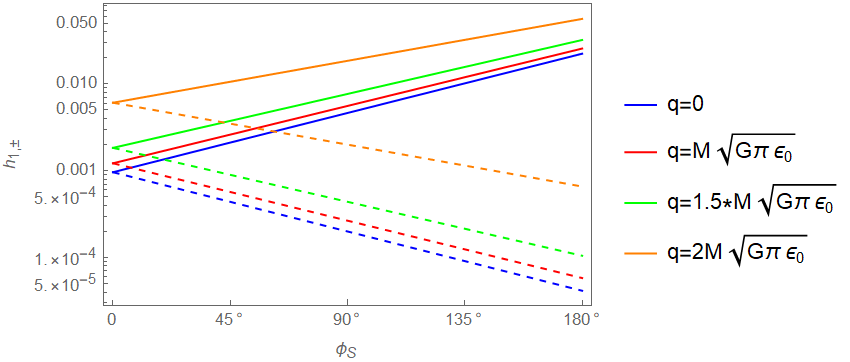}
\caption{Height of the step for the first order images ($n=1$) for the RN metric with different values of the electric charge $q$. Solid lines represent the heights for the steps due to the positive parity image while dashed lines are for negative parity images.}
\label{AltezzaRN}
\end{figure}

\begin{figure}
\centering
\includegraphics[width=\linewidth]{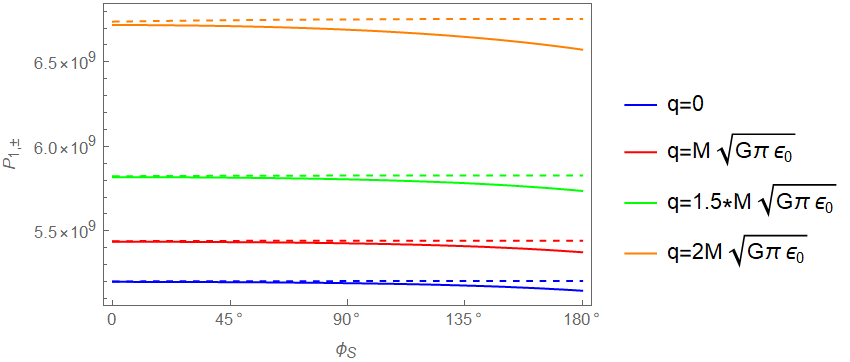}
\caption{Periodicity in the step for the first order images ($n=1$) for the RN metric with different values of the electric charge $q$. Solid lines represent the periodicities for the positive parity image while dashed lines are for negative parity images.}
\label{PeriodoRN}
\end{figure}

The second metric is the Janis-Newman-Winicour (JNW) metric \cite{Janis1968}, generated by the interaction of the metric tensor with an additional scalar field

\begin{equation}
A(r)=c^2\left(1-\frac{2 M G}{c^2 r}\right)^\gamma \qquad B(r)=\left(1-\frac{2 M G}{c^2 r}\right)^{-\gamma} \qquad C(r)=\left(1-\frac{2 M G}{c^2 r}\right)^{1-\gamma}r^2.
\end{equation}

\begin{figure}
\centering
\includegraphics[width=\linewidth]{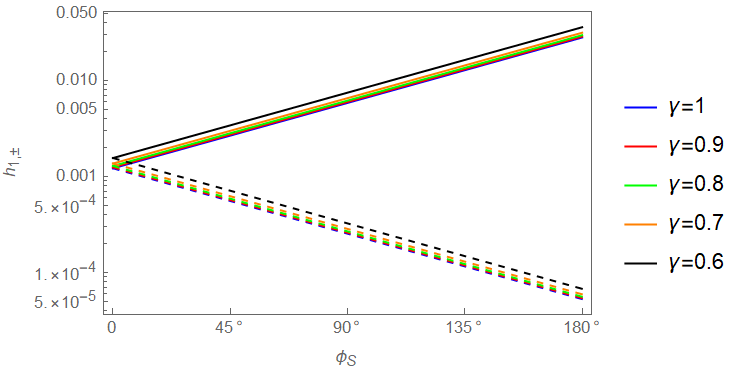}
\caption{Height of the step for the first order images ($n=1$) for the JNW metric with different values of the parameter $\gamma$ related to the scalar charge. Solid lines represent the heights for the steps due to the positive parity image while dashed lines are for negative parity images.}
\label{AltezzaJNW}
\end{figure}

\begin{figure}
\centering
\includegraphics[width=\linewidth]{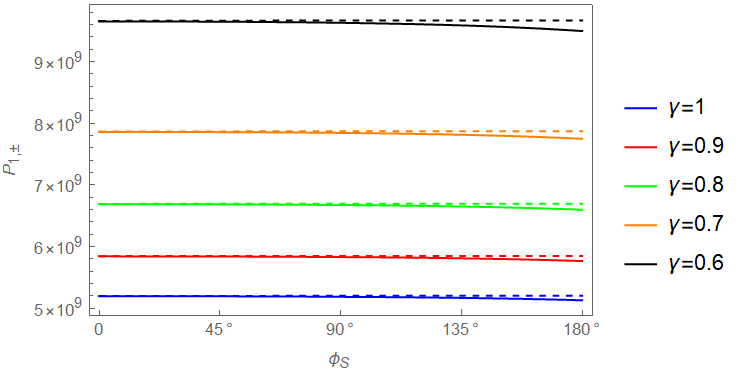}
\caption{Periodicity in the step for the first order images ($n=1$) for the JNW metric with different values of the parameter $\gamma$ related to the scalar charge. Solid lines represent the periodicities for the positive parity image while dashed lines are for negative parity images.}
\label{PeriodoJNW}
\end{figure}

The RN spacetime reduces to Schwarzschild for vanishing electric charge $q$. The JNW metric depends on the scalar charge through $\gamma=M/\sqrt{M^2+q^2}$. We thus have $0<\gamma \leq 1$, recovering Schwarzschild for $\gamma=1$.
On the other hand, a RN metric with $q>2M\sqrt{G \epsilon_0 \pi}$  has no event horizon and causality violations appear \cite{hawking1973large}. Nevertheless, a photon sphere still exists up to $q=3/\sqrt{2} \; M\sqrt{G \epsilon_0 \pi}$. In JNW we have a photon sphere for $\gamma>\frac{1}{2}$ \cite{virbhadra2002gravitational}.

Figs. \ref{AltezzaRN} shows the heights of the steps in the visibility function for first order images with positive or negative parity as functions of $\phi_S$. Different colors are for different values of the electric charge $q$. As already mentioned, the steps due to positive and negative parity images of the same order coincide for perfectly aligned sources $\phi_S=0$. As the source moves out of perfect alignment, the positive parity step grows while the negative parity step decreases its height. As we turn the electric charge on, both heights increase and the slope of the lines softens. According to Eq. (\ref{aequation}), the coefficient $a$ can be obtained by the ratio $h_{1,+}/h_{2,+}$. In RN metric, this coefficient slightly increases with the electric charge \cite{eiroa2002reissner,bozza2002gravitational}. The ability to distinguish a Schwarzschild black hole from an electrically charged black hole depends on the precision at which we can measure this ratio.

The periods in the oscillations, shown in Fig. \ref{PeriodoRN}, increase with the electric charge, mostly because of the decrease in the radius of the shadow $\theta_m$. From these periods it is possible to extract the other coefficient $b$ through Eq. (\ref{bequation}).

Fig. \ref{AltezzaJNW} shows a similar behavior for the heights of the steps in the JNW metric. These increase as we move away from the pure Schwarzschild case. The increase in the periods in Fig. \ref{PeriodoJNW} looks however quite different from that of RN. So, in principle, it should be possible to use the information stored in the heights and periods of the steps to distinguish between different metrics.

\section{Orbital motion and time delay}
\label{Sec alignment}

In the previous sections we have derived the configuration of the images and their complex visibility under the assumption of a static source at fixed coordinates $(r_S,\vartheta_S=\pi/2,\phi_S)$. In this situation, the images seen by the observer lie on the line joining the projections of the source and the lens on the observer sky. Thanks to this alignment, the complex visibility has a non-trivial structure only along the $u$-axis and can be described by simple analytical formulae (\ref{visibilitysimple}). However, real sources typically orbit around the supermassive black hole at very high speeds, becoming even relativistic in the inner regions of the accretion disk. Different images created by photons emitted by such sources are subject to time delays depending on the lengths of the different paths taken by the photons and on the gravitational fields crossed along their journeys.

We may thus ask how slow should a compact source be to generate a staircase pattern as those studied in this paper. To have an idea, we consider a source on a circular orbit with radius $r_S$ moving along the meridional direction $\vartheta$, so as to maximize the misalignment of the higher order images. The orbital period of such source, as measured by the observer at infinity, would be
\begin{equation}
    P_S=2\pi\sqrt{\frac{C'(r_S)}{A'(r_S)}}=2\pi \frac{r_S^{3/2}}{\sqrt{GM}},
\end{equation}
where in the last equality we have specified the functions $C$ and $A$ to the Schwarzschild metric and we have thus recovered the third Kepler's law.

The time delay between one image of order $n$ and the next image of order $n+1$ is essentially given by the time taken by a light ray to complete one turn around the photon sphere \cite{BozzaMancini2004,Hadar2021}
\begin{equation}
    \Delta T=2\pi\sqrt{\frac{C'(r_m)}{A'(r_m)}}=3\sqrt{3}\pi \frac{GM}{c^3}.
\end{equation}

The angular displacement of these two images of consecutive order coincides with the angular displacement of the source due to its orbital motion in the time $\Delta T$:
\begin{equation}
    \Delta \vartheta = \frac{2\pi}{\sin \phi_S} \frac{\Delta T}{P_S}=  \frac{2\pi}{\sin \phi_S} \left( \frac{3GM}{c^2 r_S} \right)^{3/2}.
\end{equation}

We can quantify the degree of misalignment of the images by taking the ratio between the angular displacement and the angular extension of the images:
\begin{equation}
\mu\equiv \frac{\Delta \vartheta}{\Delta \vartheta_S}.
\end{equation}

If we assume a spherical source, we may set $\Delta \vartheta_S = \Delta r_S/(r_S \sin \phi_S)$, where $\Delta r_S$ is the physical size of the source. We would then have
\begin{equation}
    \mu = \frac{2\pi}{r_S^{1/2} \Delta r_S} \left( \frac{3GM}{c^2} \right)^{3/2}.
\end{equation}

\begin{figure}
\includegraphics[width=17cm, right]{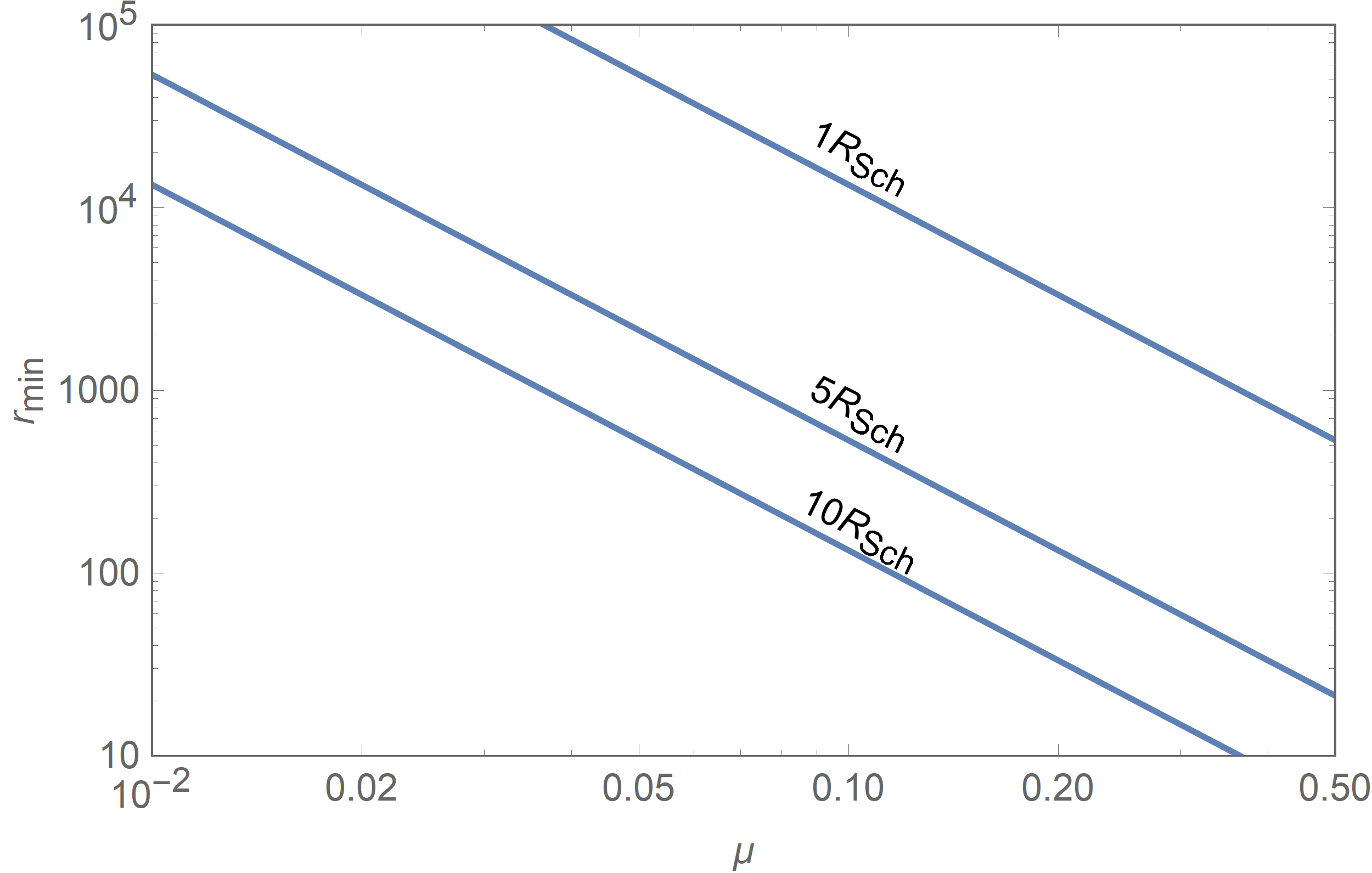}
\caption{Minimum distance (in Schwarzschild radii) as a function of the misalignment $\mu$ between images of consecutive order for sources of the indicated size, under the hypothesis of circular orbital motion.}
\label{Fig threshold}
\end{figure}

For a gaussian-shaped source, even a misalignment of $10\%$ would be quite acceptable and the scheme proposed in this paper would work fine. However, if the source has a size of the order of the Schwarzschild radius, the misalignment would anyway scale as $\left(R_{Sch}/r_S\right)^{1/2}$, which means that we need a source farther than 100 Schwarzschild radii to have a misalignment lower than $10\%$. Fig. \ref{Fig threshold} shows the minimum distance for sources of various sizes as a function of the corresponding misalignment. This shows that indeed we may have well-aligned images described by the visibility (\ref{visibilitysimple}), provided the source is far enough.

\section{Discussion and conclusions}
\label{Conclusion}

This paper provides a modern follow-up to Refs. \cite{bozza2002gravitational} and \cite{bozza2007strong} in the light of the EHT observations \cite{akiyama2019first1}. The strong deflection limit formalism allows a complete analytical derivation of the shape of the complex visibility function for a compact source at some distance from a black hole or any ultra-compact objects endowed with a photon sphere. We have been able to track the height and width of the steps generated by higher order images, along with the periodicities in the angular frequency, to the coefficients of the strong deflection limit expansion. These formulae establish a direct connection between the features in the visibility and the form of the spacetime metric that shapes the photon sphere. 

The application of such ideas to a real astrophysical case obviously requires space-based interferometry, with different challenges for different spectral bands.

By operating at mm wavelength the EHT is sensitive to non-thermal radiation from accreting matter. A compact source may arise from very energetic flare, i.e. a localized change in the magnetic fields surrounding the black hole triggering the emission of radiation from a relatively small region \cite{EHTpolarizationVIII2021}. The advantage of a compact source is that all higher order images are small and well-separated, allowing an easy disentangling between the properties of the metric and the source details. In fact, we have encoded the source effective size in a single parameter $\Delta\theta_0$, which is estimated by the steps width. In contrast, the observation of the shadow profile and the light ring features always requires some minimal hypothesis on the nature of the accretion flow and thus retains some model-dependence to some extent \cite{Chael2021}.

The detection of higher order steps would be possible with stations in the Moon or in the Lagrangian point L2 \cite{johnson2020universal,Roelofs2019}. The $v$-axis contains information on the tangential extension of the source and its state of motion. Focusing on the more interesting $u$-axis, continuous observations by stations spaced on a single direction should be able to catch signals due to sources aligned along the same direction. A small network of stations on the Moon should be able to estimate the periodicity in the visibility (\ref{Period+})-(\ref{Period-}). This is important to constrain the spatial distribution of the images, which depends on $\theta_m$ and the coefficient $b$. As discussed in Section \ref{Sec alignment}, our formalism is limited to sources that are slow (distant) enough to have images aligned on the same direction. An extension to faster (closer) sources is certainly possible by incorporating time delay in our formulae. We note that also for rotating black holes the images no longer occur on a straight line and even new images may be generated by extended caustics \cite{Bozza2005,Bozza2008}. In this case, it would be necessary to extend the network over different directions so as to have a full map of the images. In this respect, the study presented in this paper can be the basis for further generalizations to axially symmetric black holes and to fast-orbiting sources down to the event horizon. 

Interferometry at shorter wavelengths has been suggested in the context of the search for extrasolar planets \cite{DARWIN,LIFE2021}. In the mid-infrared (MIR), at $\lambda_{eff}=10 \mu m$, it would be possible to see cool giant stars with expanded atmospheres that might be orbiting the black hole at some distance \cite{Tal2017}. Such relatively compact objects would provide ideal sources for the observation of higher order images. At such wavelengths, the baseline needed to see the staircase structure in the visibility is reduced by a factor 100 with respect to mm observations. For example, the resolution reached at $\lambda=1 mm$ with a station at L2 would be obtained at $\lambda=10\mu m$ with a baseline just wider than the Earth diameter. Indeed, these projects for space interferometry in MIR will require further studies to overcome the technical challenges. Achieving stability in this band on such long distances is beyond current capabilities, but the existence of multiple interests from different fields will certainly foster new ideas in the next decades.


\bibliographystyle{IEEEtran}
\bibliography{bibliografia}

\begin{thebibliography}{10}
\providecommand{\url}[1]{#1}
\csname url@samestyle\endcsname
\providecommand{\newblock}{\relax}
\providecommand{\bibinfo}[2]{#2}
\providecommand{\BIBentrySTDinterwordspacing}{\spaceskip=0pt\relax}
\providecommand{\BIBentryALTinterwordstretchfactor}{4}
\providecommand{\BIBentryALTinterwordspacing}{\spaceskip=\fontdimen2\font plus
\BIBentryALTinterwordstretchfactor\fontdimen3\font minus
  \fontdimen4\font\relax}
\providecommand{\BIBforeignlanguage}[2]{{%
\expandafter\ifx\csname l@#1\endcsname\relax
\typeout{** WARNING: IEEEtran.bst: No hyphenation pattern has been}%
\typeout{** loaded for the language `#1'. Using the pattern for}%
\typeout{** the default language instead.}%
\else
\language=\csname l@#1\endcsname
\fi
#2}}
\providecommand{\BIBdecl}{\relax}
\BIBdecl

\bibitem{akiyama2019first1}
{Event Horizon Telescope Collaboration}, K.~Akiyama, A.~Alberdi, W.~Alef
  \emph{et~al.}, ``{First M87 Event Horizon Telescope Results. i. The Shadow of
  the Supermassive Black Hole},'' \emph{\apjl}, vol. 875, no.~1, p.~L1, Apr.
  2019.

\bibitem{akiyama2019first2}
------, ``First m87 event horizon telescope results. ii. array and
  instrumentation,'' \emph{The Astrophysical Journal Letters}, vol. 875, no.~1,
  p.~L2, 2019.

\bibitem{akiyama2019first3}
------, ``First m87 event horizon telescope results. iii. data processing and
  calibration,'' \emph{The Astrophysical Journal Letters}, vol. 875, no.~1,
  p.~L3, 2019.

\bibitem{akiyama2019first4}
------, ``First m87 event horizon telescope results. iv. imaging the central
  supermassive black hole,'' \emph{The Astrophysical Journal Letters}, vol.
  875, no.~1, p.~L4, 2019.

\bibitem{akiyama2019first5}
------, ``First m87 event horizon telescope results. v. physical origin of the
  asymmetric ring,'' \emph{The Astrophysical Journal Letters}, vol. 875, no.~1,
  p.~L5, 2019.

\bibitem{akiyama2019first6}
------, ``First m87 event horizon telescope results. vi. the shadow and mass of
  the central black hole,'' \emph{The Astrophysical Journal Letters}, vol. 875,
  no.~1, p.~L6, 2019.

\bibitem{Bardeen1973}
J.~M. {Bardeen}, ``{Timelike and null geodesics in the Kerr metric.}'' in
  \emph{Black Holes (Les Astres Occlus)}, Jan. 1973, pp. 215--239.

\bibitem{Perlick2021}
V.~{Perlick} and O.~Y. {Tsupko}, ``{Calculating black hole shadows: review of
  analytical studies},'' \emph{arXiv e-prints}, p. arXiv:2105.07101, May 2021.

\bibitem{Zhu2019}
Z.~{Zhu}, M.~D. {Johnson}, and R.~{Narayan}, ``{Testing General Relativity with
  the Black Hole Shadow Size and Asymmetry of Sagittarius A*: Limitations from
  Interstellar Scattering},'' \emph{\apj}, vol. 870, no.~1, p.~6, Jan. 2019.

\bibitem{Issaoun2019}
S.~{Issaoun}, M.~D. {Johnson}, L.~{Blackburn} \emph{et~al.}, ``{The Size,
  Shape, and Scattering of Sagittarius A* at 86 GHz: First VLBI with ALMA},''
  \emph{\apj}, vol. 871, no.~1, p.~30, Jan. 2019.

\bibitem{Zakharov2014}
A.~F. {Zakharov}, ``{Constraints on a charge in the Reissner-Nordstr{\"o}m
  metric for the black hole at the Galactic Center},'' \emph{\prd}, vol.~90,
  no.~6, p. 062007, Sep. 2014.

\bibitem{Tsukamoto2014}
N.~{Tsukamoto}, Z.~{Li}, and C.~{Bambi}, ``{Constraining the spin and the
  deformation parameters from the black hole shadow},'' \emph{\jcap}, vol.
  2014, no.~6, p. 043, Jun. 2014.

\bibitem{Bambi2019}
C.~{Bambi}, K.~{Freese}, S.~{Vagnozzi}, and L.~{Visinelli}, ``{Testing the
  rotational nature of the supermassive object M87* from the circularity and
  size of its first image},'' \emph{\prd}, vol. 100, no.~4, p. 044057, Aug.
  2019.

\bibitem{Vagnozzi2019}
S.~{Vagnozzi} and L.~{Visinelli}, ``{Hunting for extra dimensions in the shadow
  of M87*},'' \emph{\prd}, vol. 100, no.~2, p. 024020, Jul. 2019.

\bibitem{Khodadi2020}
M.~{Khodadi}, A.~{Allahyari}, S.~{Vagnozzi}, and D.~F. {Mota}, ``{Black holes
  with scalar hair in light of the Event Horizon Telescope},'' \emph{\jcap},
  vol. 2020, no.~9, p. 026, Sep. 2020.

\bibitem{Gralla2020}
S.~E. {Gralla}, A.~{Lupsasca}, and D.~P. {Marrone}, ``{The shape of the black
  hole photon ring: A precise test of strong-field general relativity},''
  \emph{\prd}, vol. 102, no.~12, p. 124004, Dec. 2020.

\bibitem{Gralla2020b}
S.~E. {Gralla} and A.~{Lupsasca}, ``{Observable shape of black hole photon
  rings},'' \emph{\prd}, vol. 102, no.~12, p. 124003, Dec. 2020.

\bibitem{Farah2020}
J.~R. {Farah}, D.~W. {Pesce}, M.~D. {Johnson}, and L.~{Blackburn}, ``{On the
  Approximation of the Black Hole Shadow with a Simple Polar Curve},''
  \emph{\apj}, vol. 900, no.~1, p.~77, Sep. 2020.

\bibitem{Shaikh2021}
R.~{Shaikh}, K.~{Pal}, K.~{Pal}, and T.~{Sarkar}, ``{Constraining alternatives
  to the Kerr black hole},'' \emph{\mnras}, Jun. 2021.

\bibitem{Luminet1979}
J.~P. {Luminet}, ``{Image of a spherical black hole with thin accretion
  disk.}'' \emph{\aap}, vol.~75, pp. 228--235, May 1979.

\bibitem{johnson2020universal}
M.~D. {Johnson}, A.~{Lupsasca}, A.~{Strominger}, G.~N. {Wong} \emph{et~al.},
  ``{Universal interferometric signatures of a black hole's photon ring},''
  \emph{Science Advances}, vol.~6, no.~12, p. eaaz1310, Mar. 2020.

\bibitem{Origins2019}
D.~{Pesce}, K.~{Haworth}, G.~J. {Melnick} \emph{et~al.}, ``{Extremely long
  baseline interferometry with Origins Space Telescope},'' in \emph{Bulletin of
  the American Astronomical Society}, vol.~51, Sep. 2019, p. 176.

\bibitem{Radioastron2021}
A.~S. {Andrianov}, A.~M. {Baryshev}, H.~{Falcke} \emph{et~al.}, ``{Simulations
  of M87 and Sgr A* imaging with the Millimetron Space Observatory on
  near-Earth orbits},'' \emph{\mnras}, vol. 500, no.~4, pp. 4866--4877, Jan.
  2021.

\bibitem{Dokuchaev2019}
V.~I. {Dokuchaev} and N.~O. {Nazarova}, ``{Event Horizon Image within Black
  Hole Shadow},'' \emph{Soviet Journal of Experimental and Theoretical
  Physics}, vol. 128, no.~4, pp. 578--585, Apr. 2019.

\bibitem{Dokuchaev2020a}
------, ``{Visible Shapes of Black Holes M87* and SgrA*},'' \emph{Universe},
  vol.~6, no.~9, p. 154, Sep. 2020.

\bibitem{Dokuchaev2020b}
------, ``{Silhouettes of invisible black holes},'' \emph{Physics Uspekhi},
  vol.~63, no.~6, pp. 583--600, Sep. 2020.

\bibitem{Chael2021}
A.~{Chael}, M.~D. {Johnson}, and A.~{Lupsasca}, ``{Observing the Inner Shadow
  of a Black Hole: A Direct View of the Event Horizon},'' \emph{arXiv
  e-prints}, p. arXiv:2106.00683, Jun. 2021.

\bibitem{BozzaMancini2004}
V.~{Bozza} and L.~{Mancini}, ``{Time Delay in Black Hole Gravitational Lensing
  as a Distance Estimator},'' \emph{General Relativity and Gravitation},
  vol.~36, no.~2, pp. 435--450, Feb. 2004.

\bibitem{Chesler2020}
P.~M. {Chesler}, L.~{Blackburn}, S.~S. {Doeleman} \emph{et~al.}, ``{Light echos
  and coherent autocorrelations in a black hole spacetime},'' \emph{arXiv
  e-prints}, p. arXiv:2012.11778, Dec. 2020.

\bibitem{Hadar2021}
S.~{Hadar}, M.~D. {Johnson}, A.~{Lupsasca}, and G.~N. {Wong}, ``{Photon ring
  autocorrelations},'' \emph{\prd}, vol. 103, no.~10, p. 104038, May 2021.

\bibitem{Wong2021}
G.~N. {Wong}, ``{Black Hole Glimmer Signatures of Mass, Spin, and
  Inclination},'' \emph{\apj}, vol. 909, no.~2, p. 217, Mar. 2021.

\bibitem{EHTpolarizationVII2021}
{Event Horizon Telescope Collaboration}, K.~{Akiyama}, J.~C. {Algaba},
  A.~{Alberdi} \emph{et~al.}, ``{First M87 Event Horizon Telescope Results.
  VII. Polarization of the Ring},'' \emph{\apjl}, vol. 910, no.~1, p. L12, Mar.
  2021.

\bibitem{EHTpolarizationVIII2021}
------, ``{First M87 Event Horizon Telescope Results. VIII. Magnetic Field
  Structure near The Event Horizon},'' \emph{\apjl}, vol. 910, no.~1, p. L13,
  Mar. 2021.

\bibitem{Himwich2020}
E.~{Himwich}, M.~D. {Johnson}, A.~{Lupsasca}, and A.~{Strominger}, ``{Universal
  polarimetric signatures of the black hole photon ring},'' \emph{\prd}, vol.
  101, no.~8, p. 084020, Apr. 2020.

\bibitem{Moscibrodzka2021}
M.~{Moscibrodzka}, A.~{Janiuk}, and M.~{De Laurentis}, ``{Unraveling circular
  polarimetric images of magnetically arrested accretion flows near event
  horizon of a black hole},'' \emph{arXiv e-prints}, p. arXiv:2103.00267, Feb.
  2021.

\bibitem{Tsupko2020}
O.~Y. {Tsupko}, Z.~{Fan}, and G.~S. {Bisnovatyi-Kogan}, ``{Black hole shadow as
  a standard ruler in cosmology},'' \emph{Classical and Quantum Gravity},
  vol.~37, no.~6, p. 065016, Mar. 2020.

\bibitem{Vagnozzi2020}
S.~{Vagnozzi}, C.~{Bambi}, and L.~{Visinelli}, ``{Concerns regarding the use of
  black hole shadows as standard rulers},'' \emph{Classical and Quantum
  Gravity}, vol.~37, no.~8, p. 087001, Apr. 2020.

\bibitem{darwin1959gravity}
C.~G. {Darwin}, ``The gravity field of a particle,'' \emph{Proceedings of the
  Royal Society of London. Series A. Mathematical and Physical Sciences}, vol.
  249, no. 1257, pp. 180--194, 1959.

\bibitem{Ohanian1987}
H.~C. {Ohanian}, ``{The black hole as a gravitational ``lens''},''
  \emph{American Journal of Physics}, vol.~55, no.~5, pp. 428--432, May 1987.

\bibitem{virbhadra2000schwarzschild}
K.~S. {Virbhadra} and G.~F.~R. {Ellis}, ``{Schwarzschild black hole lensing},''
  \emph{\prd}, vol.~62, no.~8, p. 084003, Oct. 2000.

\bibitem{bozza2002gravitational}
V.~{Bozza}, ``{Gravitational lensing in the strong field limit},'' \emph{\prd},
  vol.~66, no.~10, p. 103001, Nov. 2002.

\bibitem{Bozza2010}
------, ``{Gravitational lensing by black holes},'' \emph{General Relativity
  and Gravitation}, vol.~42, no.~9, pp. 2269--2300, Sep. 2010.

\bibitem{Cardoso2009}
\BIBentryALTinterwordspacing
V.~Cardoso, A.~S. Miranda, E.~Berti, H.~Witek, and V.~T. Zanchin, ``Geodesic
  stability, lyapunov exponents, and quasinormal modes,'' \emph{Phys. Rev. D},
  vol.~79, p. 064016, Mar 2009. [Online]. Available:
  \url{https://link.aps.org/doi/10.1103/PhysRevD.79.064016}
\BIBentrySTDinterwordspacing

\bibitem{stefanov2010connection}
I.~Z. {Stefanov}, S.~S. {Yazadjiev}, and G.~G. {Gyulchev}, ``{Connection
  between Black-Hole Quasinormal Modes and Lensing in the Strong Deflection
  Limit},'' \emph{\prl}, vol. 104, no.~25, p. 251103, Jun. 2010.

\bibitem{wei2014establishing}
S.-W. {Wei} and Y.-X. {Liu}, ``{Establishing a universal relation between
  gravitational waves and black hole lensing},'' \emph{\prd}, vol.~89, no.~4,
  p. 047502, Feb. 2014.

\bibitem{raffaelli2016strong}
B.~{Raffaelli}, ``{Strong gravitational lensing and black hole quasinormal
  modes: towards a semiclassical unified description},'' \emph{General
  Relativity and Gravitation}, vol.~48, p.~16, Feb. 2016.

\bibitem{Claudel2001}
C.-M. {Claudel}, K.~S. {Virbhadra}, and G.~F.~R. {Ellis}, ``{The geometry of
  photon surfaces},'' \emph{Journal of Mathematical Physics}, vol.~42, no.~2,
  pp. 818--838, Feb. 2001.

\bibitem{Cunha2017}
P.~V.~P. {Cunha}, E.~{Berti}, and C.~A.~R. {Herdeiro}, ``{Light-Ring Stability
  for Ultracompact Objects},'' \emph{\prl}, vol. 119, no.~25, p. 251102, Dec.
  2017.

\bibitem{chandrasekhar1998mathematical}
S.~{Chandrasekhar}, \emph{{The Mathematical Theory of Black Holes}}, 1998.

\bibitem{Teo2003}
E.~{Teo}, ``{Spherical Photon Orbits Around a Kerr Black Hole},'' \emph{General
  Relativity and Gravitation}, vol.~35, no.~11, pp. 1909--1926, Nov. 2003.

\bibitem{bozza2007strong}
V.~Bozza and G.~Scarpetta, ``Strong deflection limit of black hole
  gravitational lensing with arbitrary source distances,'' \emph{Physical
  Review D}, vol.~76, no.~8, p. 083008, 2007.

\bibitem{falcke1999viewing}
H.~{Falcke}, F.~{Melia}, and E.~{Agol}, ``{Viewing the Shadow of the Black Hole
  at the Galactic Center},'' \emph{\apjl}, vol. 528, no.~1, pp. L13--L16, Jan.
  2000.

\bibitem{Gralla2019}
S.~E. {Gralla}, D.~E. {Holz}, and R.~M. {Wald}, ``{Black hole shadows, photon
  rings, and lensing rings},'' \emph{\prd}, vol. 100, no.~2, p. 024018, Jul.
  2019.

\bibitem{SEF}
P.~{Schneider}, J.~{Ehlers}, and E.~E. {Falco}, \emph{{Gravitational Lenses}},
  1992.

\bibitem{Bozza2005}
V.~{Bozza}, F.~{de Luca}, G.~{Scarpetta}, and M.~{Sereno}, ``{Analytic Kerr
  black hole lensing for equatorial observers in the strong deflection
  limit},'' \emph{\prd}, vol.~72, no.~8, p. 083003, Oct. 2005.

\bibitem{Mashoon1985}
\BIBentryALTinterwordspacing
B.~Mashhoon, ``Stability of charged rotating black holes in the eikonal
  approximation,'' \emph{Phys. Rev. D}, vol.~31, pp. 290--293, Jan 1985.
  [Online]. Available: \url{https://link.aps.org/doi/10.1103/PhysRevD.31.290}
\BIBentrySTDinterwordspacing

\bibitem{Janis1968}
A.~I. {Janis}, E.~T. {Newman}, and J.~{Winicour}, ``{Reality of the
  Schwarzschild Singularity},'' \emph{\prl}, vol.~20, no.~16, pp. 878--880,
  Apr. 1968.

\bibitem{hawking1973large}
S.~W. Hawking and G.~F.~R. Ellis, \emph{The large scale structure of
  space-time}.\hskip 1em plus 0.5em minus 0.4em\relax Cambridge university
  press, 1973, vol.~1.

\bibitem{virbhadra2002gravitational}
K.~{Virbhadra} and G.~{Ellis}, ``Gravitational lensing by naked
  singularities,'' \emph{Physical Review D}, vol.~65, no.~10, p. 103004, 2002.

\bibitem{eiroa2002reissner}
E.~F. {Eiroa}, G.~E. {Romero}, and D.~F. {Torres}, ``{Reissner-Nordstr{\"o}m
  black hole lensing},'' \emph{\prd}, vol.~66, no.~2, p. 024010, Jul. 2002.

\bibitem{Roelofs2019}
F.~{Roelofs}, H.~{Falcke}, C.~{Brinkerink}, M.~{Mo{\'s}cibrodzka}, L.~I.
  {Gurvits}, M.~{Martin-Neira}, V.~{Kudriashov}, M.~{Klein-Wolt}, R.~{Tilanus},
  M.~{Kramer}, and L.~{Rezzolla}, ``{Simulations of imaging the event horizon
  of Sagittarius A* from space},'' \emph{\aap}, vol. 625, p. A124, May 2019.

\bibitem{Bozza2008}
V.~{Bozza}, ``{Optical caustics of Kerr spacetime: The full structure},''
  \emph{\prd}, vol.~78, no.~6, p. 063014, Sep. 2008.

\bibitem{DARWIN}
A.~{Leger} and T.~{Herbst}, ``{DARWIN mission proposal to ESA},'' \emph{arXiv
  e-prints}, p. arXiv:0707.3385, Jul. 2007.

\bibitem{LIFE2021}
S.~P. {Quanz}, M.~{Ottiger}, E.~{Fontanet}, J.~{Kammerer} \emph{et~al.},
  ``{Large Interferometer For Exoplanets (LIFE): I. Improved exoplanet
  detection yield estimates for a large mid-infrared space-interferometer
  mission},'' \emph{arXiv e-prints}, p. arXiv:2101.07500, Jan. 2021.

\bibitem{Tal2017}
T.~{Alexander}, ``{Stellar Dynamics and Stellar Phenomena Near a Massive Black
  Hole},'' \emph{\araa}, vol.~55, no.~1, pp. 17--57, Aug. 2017.

\end{thebibliography}
\end{document}